# Augmenting the availability of historical GDP per capita estimates through machine learning

Philipp Koch[1,2,*], Viktor Stojkoski[1,3], César A. Hidalgo[1,4,5,*].

[1] Center for Collective Learning, ANITI, IRIT, Université de Toulouse, Toulouse, France.
[2] EcoAustria – Institute for Economic Research, Vienna, Austria.
[3] Faculty of Economics, University Ss. Cyril and Methodius, Skopje, North Macedonia.
[4] Center for Collective Learning, CIAS, Corvinus University, Budapest, Hungary.
[5] Toulouse School of Economics, Université de Toulouse, Toulouse, France.

*Corresponding authors: Philipp Koch (philipp.koch@ecoaustria.ac.at) and César A. Hidalgo (cesar.hidalgo@tse-fr.eu)

**Abstract**

Can we use data on the biographies of historical figures to estimate the GDP per capita of countries and regions? Here we introduce a machine learning method to estimate the GDP per capita of dozens of countries and hundreds of regions in Europe and North America for the past 700 years starting from data on the places of birth, death, and occupations of hundreds of thousands of historical figures. We build an elastic net regression model to perform feature selection and generate out-of-sample estimates that explain 90% of the variance in known historical income levels. We use this model to generate GDP per capita estimates for countries, regions, and time periods for which this data is not available and externally validate our estimates by comparing them with four proxies of economic output: urbanization rates in the past 500 years, body height in the 18th century, wellbeing in 1850, and church building activity in the 14th and 15th century. Additionally, we show our estimates reproduce the well-known reversal of fortune between southwestern and northwestern Europe between 1300 and 1800 and find this is largely driven by countries and regions engaged in Atlantic trade. These findings validate the use of fine-grained biographical data as a method to produce historical GDP per capita estimates. We publish our estimates with confidence intervals together with all collected source data in a comprehensive dataset.

**Significance Statement**

The scarcity of historical GDP per capita limits our ability to explore questions of long-term economic growth and development. Here, we introduce a machine learning method to estimate the GDP per capita of dozens of countries and hundreds of regions in Europe and North America for the past 700 years using data on hundreds of thousands of historical figures. We find that our model generates accurate out-of-sample estimates ($R^2$=90%) that correlate with other proxies of economic output such as urbanization, body height, wellbeing, and church building activity. These findings showcase the use of granular biographical data to estimate historical income levels and augment the availability of historical GDP per capita data.

## Introduction

During the last decades, machine learning methods helped expand the economics toolbox (1, 2), from the use of satellite images to estimate poverty (3–6), population (7, 8), and land use (9–12), to the use of recommender systems to support economic diversification policies (13–16). But machine learning methods are not only useful to study the present or predict the future, they can also be used to explore the past. In this paper, we introduce a machine learning method designed to reconstruct historical GDP per capita estimates of dozens of European and North American countries and regions for the past 700 years, more than quadrupling the availability of historical economic output data for these regions.

For decades, economic historians have made great efforts to reconstruct the GDP per capita of countries and regions using historical documents on economic output (17, 18), and by approximating GDP per capita using data on consumption (19–26). Despite these efforts, estimates of historical GDPs per capita are still scarce (Figs. 1 A-B). The Maddison project, the largest collection of historical GDP per capita estimates (27, 28), has data for only 11 European countries for the year 1750 and 5 for the 1300s: France, England, Spain, Sweden, and Northern Italy. This leaves out important economies, such as those of Austria, Russia, and Switzerland in the 1750s, and those of most of Europe during the renaissance. GDP per capita estimates on a smaller geographic scale such as administrative regions or cities are even more scarce. For the year 1750, for instance, we only found regional GDPs per capita for Spain (29) and Sweden (30).

This lack of data limits our ability to explore questions of long-term economic growth and development. Yet, research on how to extend these estimates using big data and machine learning methods is still relatively unexplored. Here, we ask whether data on the biographies of hundreds of thousands of historical figures, combined with machine learning methods, can be used to extend GDP per capita estimates to countries, regions, and time periods for which this data is not available.

The use of data on historical figures is not a capricious choice. On the one hand, unlike data on GDPs per capita, there is an abundance of accurate biographical records. Recent research efforts have made available structured data on the places of birth, death, and occupations of hundreds of thousands of historical figures (31, 32), providing a potentially rich source of features that should correlate with regional variations in GDPs per capita. On the other hand, there are good reasons why the GDP per capita of a country or region should correlate with the probability that a historical figure is born or has died there.

Consider both direct and indirect channels. Inventors and scientists involved in productivity-enhancing and lifesaving innovations—such as James Watt and Alexander Fleming—may contribute directly to the GDP per capita of their economies (33–35) by increasing productivity or reducing disease burden. But there are also important indirect channels. Wealthier regions are more likely to attract talent, make local talent more visible, and provide the freedom and opportunities needed for individuals to specialize in cultural and economic activities. It is well known, for instance, that individuals that become famous—and get recorded historically—tend to be remarkably mobile (36–39). We should also expect these migratory forces to attract talented individuals to locations that are rich in terms of physical and human capital (39–48). For the sake of generating historical estimates of economic development, we are indifferent about whether wealth attracts talent, whether wealth makes talent more visible, or whether talent contributes directly to wealth. All of these channels imply a positive correlation with wealth that should be mineable from biographical records. In fact, our estimates do not require us to identify a causal link between any of these channels and GDP per capita, but to identify robust correlations between the presence of historical figures and the GDP per capita of the countries and regions where those individuals once located. That is, the careers of Michelangelo, Sandro Botticelli, and Filippo Lippi, tell us something about the prosperity of Tuscany in the 15$^{th}$ century, no matter whether they contributed to the wealth of Tuscany or were its by-products.

In this paper, we leverage information on more than 563K historical figures recorded across multiple languages in Wikipedia (31, 32) to test whether this data can be used to model the GDP per capita of hundreds of regions in Europe and North America for the past 700 years. Specifically, we train a set of supervised machine learning models (elastic net regression models) with geographical features derived from the biographies of famous historical figures to generate out-of-sample estimates of national and regional GDPs per capita (see Fig. 1 for a visual summary of the idea). We find the model provides encouraging results. In an out-of-sample test, it predicts the GDP per capita of European and North American countries and regions with an $R^2$=90.1% and a mean absolute error of 22.6% of the GDP per capita observed during that time period.

We externally validate these estimates by recreating qualitatively well-known historical development trajectories and by comparing them with other proxies of per capita wealth. First, we recreate the established finding that England and the Low Countries experienced larger economic growth than Southern Europe between 1300 and 1800 (49–53). We find that a large share of this reversal of fortune can be attributed to the rise of Atlantic trade, supporting earlier findings by Acemoglu, Johnson & Robinson (49). Second, we show our estimates correlate with proxies of economic development, such as urbanization rates between 1500 and 1950 (54), body height in the $18^{th}$ century (55), wellbeing in 1850 (56), and church building activity in the $14^{th}$ and $15^{th}$ century (57). These findings contribute a new method for the generation of historical GDP per capita estimates and open a door to the use of structured historical data for the estimation of long-term economic time series.

## Data

### Historical GDP per capita data

Our method builds on country-level GDP per capita estimates provided by the 2020 release of the Maddison project (27, 28). These are country-level estimates considering changing geographies. For instance, Great Britain data up to 1700 refers only to England (18), and data on Italy refers only to Northern Italy up to 1861 (20) (Figs. 1 A-B). For a full list of border changes, see the Maddison project (17–26, 29) and section 1 of the SI Appendix.

We augment Maddison's country-level data with sources for estimates on the historical GDP per capita of regions (Fig. 1 B) in Spain between 1500 and 1800 (29), in Sweden between 1571 and 1950 (30, 58), in France in 1850 (59, 60), in the United Kingdom (61, 62) and Italy (63) between 1850 and 1950, and in Portugal (64) and Belgium (65) in 1900 and 1950.

Lastly, we add regional GDP per capita data for the year 2000 for most regions in the dataset. Specifically, we collect official data from Eurostat (66), the Office for National Statistics in the UK (67), the Bureau of Economic Analysis in the United States (68), Statistics Canada (69), the State Statistics Service of Ukraine (70), Belstat in Belarus (71), and Rosstat in Russia (72).

In total, we collect 1,336 GDP per capita observations in 50-year intervals (1300, 1350, …, 1950, 2000). All GDP per capita data is denoted in 2011 USD PPP, matching the unit provided in the Maddison project (SI Appendix section 1).

While the Maddison project is a comprehensive and widely used database on historical GDP per capita levels, its data must be understood as estimates. Comparing long-term economic development across the globe does not only require collecting and digitizing historical records, but also finding methods to compare purchasing powers across countries and continents. The latter are debated in the literature. For instance, it is argued that real income levels in the United States might have surpassed the ones in Europe earlier than data in the Maddison project claims (73, 74), or that the real income gap between Europe and Asia prior to the Industrial Revolution was far less pronounced (75). Despite this criticism, we use the Maddison project as gold-standard data since it has a large coverage and is still revised regularly by researchers at the University of Groningen (76).

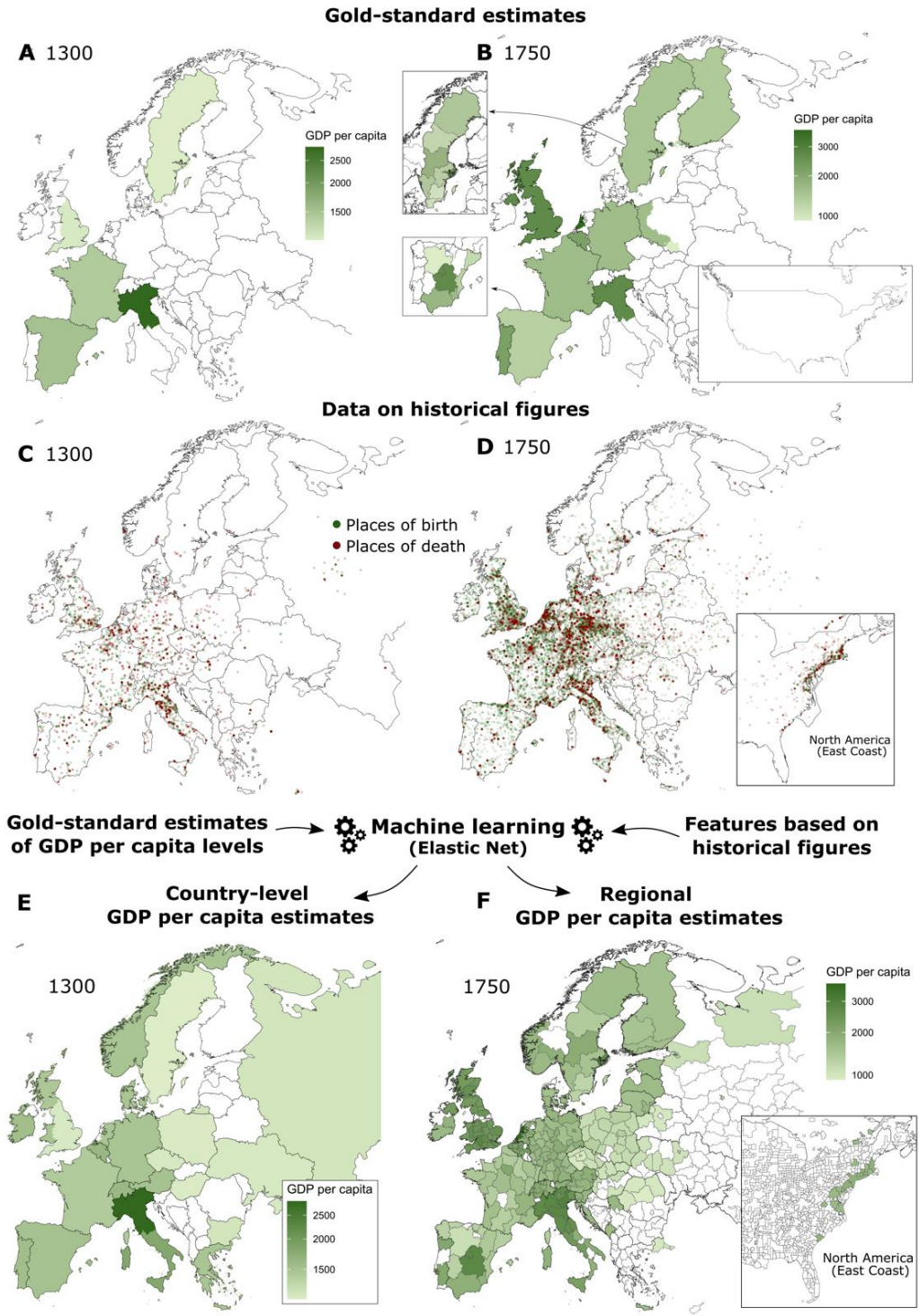

**Figure 1. Method Summary** (**A-B**) Gold-standard estimates on historical GDP per capita in (**A**) 1300 and (**B**) 1750 from the Maddison project and other sources for regional estimates (in 2011 USD). (**C-D**) Places of birth and death of famous individuals born at most 150 years prior to (**C**) 1300 and (**D**) 1750. (**E-F**) GDP per capita estimates based on available source data and machine learning models for (**E**) countries in 1300 and (**F**) regions in 1750.

**Data on historical figures**

We use data on historical figures from a recently published database of notable people recorded on Wikipedia and Wikidata, curated and cross-verified by Laouenan et al. (31). This database contains information on 2.29 million historical figures, including their places of birth, death, occupation, and proxies of their present-day popularity, such as Wikipedia page views or the number of language editions.

Data from Wikipedia is known to be subject to biases (77). For instance, famous figures of the Western world are overrepresented (78). Consider the 237K biographies Wikipedia provides of historical figures who are born between 1100 and 1900 (in at least two language editions and with an identifiable occupation). 77.9 percent of those biographies are about people who lived in Europe or North America. This is in contrast with population estimates showing that only 18.75 percent of the global population in 1820 lived in Europe or North America (27, 28). Also, cultural norms impact the portrayal of certain individuals in different language editions (79, 80), and the relative coverage of topics (81). Still, empirical studies find that the information available in Wikipedia is of relatively high accuracy when assessed by experts (82) or compared with other encyclopedias such as Britannica (83).

We address these limitations in two ways. First, we focus only on Europe and North America due to the limited representativity of other parts of the world. Second, we address potential language biases by considering only biographies with Wikipedia pages in at least two languages (to avoid including local biographies that are available only in a major language, such as English or French). We validate this methodological choice by comparing our results with models using data only from English pages or only non-English pages. We find similar results for all three samples suggesting that Wikipedia's English bias is not driving our estimates (SI Appendix section 5.5.2).

In total, we use 562,962 biographies of individuals living in Europe or North America after 1100 with an identifiable occupation and Wikipedia pages in at least two language editions (SI Appendix section 3.1). We assign biographies to countries and regions based on their places of birth and death (Fig. 1 C-D). To assign biographies to countries, we consider all border changes described in the source materials of the Maddison project (17–26, 29). For regions, we rely on European NUTS-2 regions (2021 edition), metro- and micropolitan statistical areas for the United States, metropolitan areas for Canada, and regions of similar size for other countries, e.g. oblasts in Russia (SI Appendix section 2.1). Finally, while the places of birth and death of historical figures do not provide a comprehensive view of their life history (e.g. Einstein was born in southern Germany and died in New Jersey, but lived also in Zurich and Berlin), they provide a proxy that has been used frequently in recent literature on historical migration (31, 36, 84, 85). In a recent publication (37), we tested this proxy by randomly sampling 200 individuals and manually verifying their respective Wikipedia pages, finding that in 90% of the sample it was valid (SI Appendix section 3.4).

**Feature construction**

We use this data to construct geographic features for each country, region, and time period. These include the total number of historical figures born in, died in, immigrated to (died in the place but born elsewhere), or emigrated from (born in the place but died elsewhere) each location; and occupation-specific counts (e.g. number of inventors or painters born, died, immigrated to, and emigrated from each location). These features are then weighted by an estimate of the historical popularity of each individual (the *Historical Popularity Index* ($HPI$) introduced in the Pantheon database (32)) and linearized using logarithms. $HPI$ is an estimate of historical fame breaking the barriers of space, time, and language. It combines information on the number of Wikipedia pageviews, the number of language editions, and the age of historical figures (*Materials & Methods*). For a validation of the $HPI$ see Yu et al. (32). Also, we calculate the average age of famous individuals, since increases in life expectancy have been shown to be leading indicators of the Industrial Revolution (85).

We augment this data with vectors generated using dimensionality reduction techniques such as singular value decomposition (SVD), a standard generalized eigenvalue decomposition for non-square matrices. We implement SVD by organizing our data into matrices describing the (*HPI*-weighted) number of historical figures in a location with a specific occupation. We create four different matrices for each year: births, deaths, immigrants, and emigrants, and include the first five eigenvectors of each matrix as candidate features. That is, we effectively include 20 SVD factors as potential candidate features for every year (*Materials & Methods*).

We also calculate estimates of economic complexity, an SVD type vector used frequently in economic development (16, 86, 87). The economic complexity index (*ECI*), is usually constructed with data on the geography of trade, employment, or patents, to explain cross-country and regional differences in economic growth (88–93), income inequality (92, 94) and greenhouse gas emissions (92, 95, 96). Here, we compute separate *ECI*'s for births, deaths, immigrants, and emigrants, and include them as potential features in our model (*Materials & Methods*). Finally, we include two more variables inspired by the literature on economic complexity: a location's diversity (the number of occupations with at least one individual in a location), and the average ubiquity of occupations in a location (the number of locations in which an activity, such as an occupation, is present).

Finally, there is the question of assigning features to time periods. For instance, which individuals should we consider when extracting features for the year 1600? In our model, we consider all individuals born in the 150 years prior to a respective year. That is, the features for 1600 include all biographies of individuals born between 1450 and 1600. We find our model is not too sensitive to this choice, as results using other thresholds (75, 100, and 175 years) are similar, but slightly worse than using the 150 years window (SI Appendix section 5.5.3).

In total, we collect between 250 and 300 potential features per period from the geography of famous biographies. In the next section we explain our feature selection process which is designed to avoid the risk of overfitting.

## Results

**Constructing the model**

Next, we build and validate a model of GDP per capita estimates. To avoid overfitting, we use a regularized elastic net (EN) regression model (97). Elastic net models do not simply minimize the sum of squared residuals, like an OLS regression would, but penalize the model statistics using the $\ell^1$ and $\ell^2$ norms of the coefficients, effectively performing feature selection. This allows us to identify models that provide a good predictive power with an appropriate number of features.

We should note that the selected features can be different for different time periods. Attracting painters may be a positive predictor of GDP per capita in the 16th century but not in more recent years, and begetting inventors or engineers may be correlated with economic development during the Industrial Revolution but not during the renaissance. We take this into account by selecting features separately for each period. Since limited training data renders the selection for each year impossible, we perform feature selection for five historically informed time periods within which changes in importance are less likely. Specifically, we distinguish between the Late Middle Ages (1300-1500), the Early Modern Period (1501-1750), the Age of Revolutions (1751-1850), the Machine Age (1851-1950), and the Information Age (2000). Categorizing our analysis into these distinct periods allows us to capture changing relationships between the selected features and economic development.

For each period, we train the EN model with all available source data by optimizing the hyperparameters to find the most relevant features. We optimize the model's hyperparameters using k-fold cross validation and minimizing the prediction error (*Materials & Methods*). Then, we use this model to obtain out-of-sample estimates for countries and regions in Europe and North America lacking GDP per capita data (Figs. 1 E-F). To avoid noise coming from the left-hand side of the distribution, we refrain from making predictions for locations with less than three births or deaths in a period up to 1600, with less than five births or deaths per period between 1650 and 1950, or with less than ten births or deaths in 2000. In total, we build upon our training data with 1,336 observations to provide out-of-sample estimates for 4,364 location-year combinations.

To make sure our regional estimates align with our country-level data, we rescale the regional estimates to match the population-weighted mean GDP per capita of the respective country. We use the number of births and deaths as population proxies, since data on historical population levels (54, 98) does not cover all regions in all periods and is restricted to urban population. The number of births and deaths is, however, a valid proxy of population (SI Appendix section 3.3). Lastly, we obtain standard errors and confidence intervals for our estimates by bootstrapping.

**Model performance**

We assess model performance using out-of-sample cross-validation tests and by comparing it to a baseline model. For the out-of-sample cross-validation tests, we use withheld and independent test data sets. To ensure the test data sets are independent and minimize data leakage, we remove all observations for a randomly selected 20 percent of countries, including the regions within those countries (*Materials & Methods*).

We build a baseline model that accounts for persistence in income levels and differences between supranational regions (following the United Nations geoscheme, SI Appendix section 2.2). Specifically, it is a linear regression model that predicts GDP per capita with fixed effects for supranational regions in a specific period and the GDP per capita from the end of the previous historical period. The latter variable is not available for all locations and all time periods, so we use the following hierarchical approach to fill in missing entries. First, if the GDP per capita is missing, we resort to estimates from the baseline model of the preceding period. In cases where both the source data and baseline model estimates are unavailable, we substitute with source data or estimates from the country that region is in. Finally, if none of the above is available, we use the average of the supranational region at the end of the previous period as initial GDP per capita. For example, the baseline prediction for the GDP per capita of Austria in 1800 accounts for the average GDP per capita of other Western European countries in 1800 (based on source data), as well as the GDP per capita of Austria in 1750 (based on the baseline model, since no source data is available for Austria in 1750).

The full model builds upon this baseline model and adds features derived from famous biographies. This significantly improves the predictive fit. Figs. 2 A and B are examples of how the fit improves compared to the baseline for one specific test data set consisting of Italy, Portugal, Norway, Slovenia, Albania, Croatia, Romania and Latvia. For this test data set, the fit improves from 86% (baseline model) to 89% (full model). Figs. 2 C and D show the distribution of the R-squared and the mean absolute error across 500 different randomly selected independent test sets. The fit improves, at the median, from explaining 86.2% of the variance (baseline model) to 90.1% (full model), while the mean absolute error improves from 29% of average GDP per capita to 22.6%. Kruskal-Wallis H tests on statistical differences in the distributions between the baseline model and the full model are highly significant ($p < 1e^{-15}$). We provide further details on assessing model performance in the *Materials & Methods*.

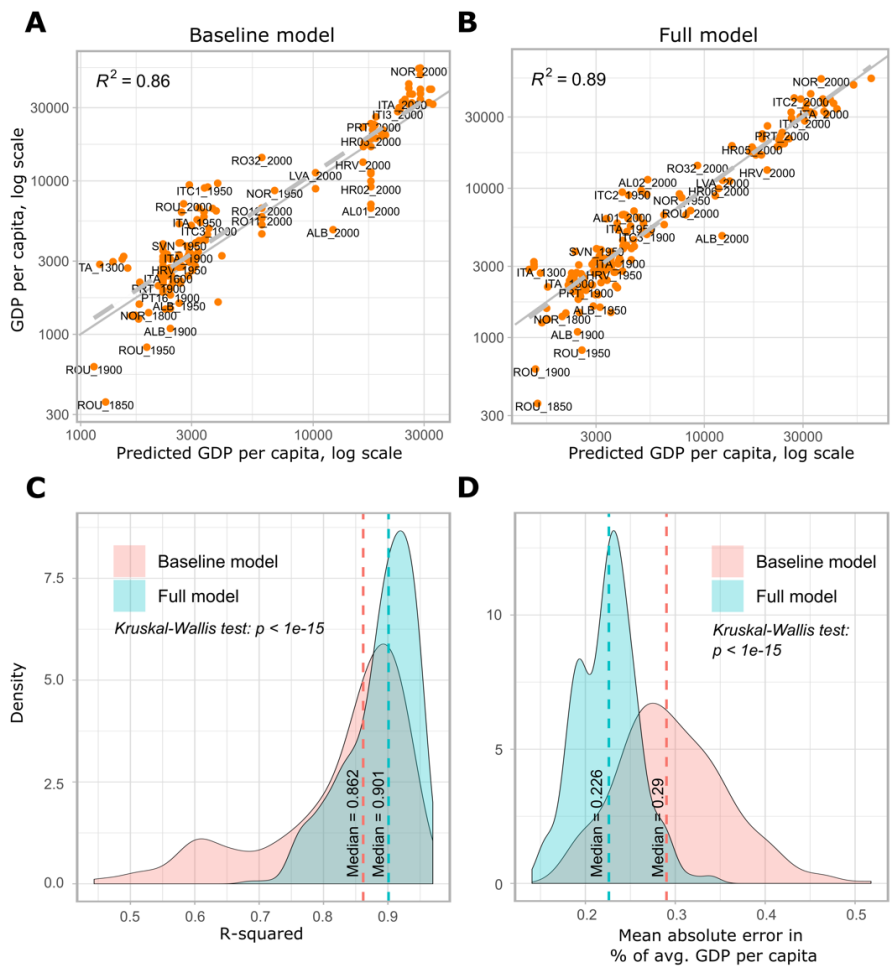

**Figure 2. Model performance.** (**A**) Baseline model prediction of test data for a random set of countries, accounting for fixed effects for supranational regions in a specific period (e.g. Southern Europe in 1950) and persistence in income levels. (**B**) Predictions of full model using elastic net. (**C**) Distribution of R-squared for the baseline and the full model when drawing 500 samples of training and test datasets. (**D**) Distribution of the mean absolute error for the baseline and the full model when drawing 500 samples of training and test datasets.

**External validation: Little Divergence, body height, wellbeing, and church building**

We externally validate our estimates in two ways.

First, we recreate Europe's well-known Little Divergence (49–53) and explore the role Atlantic trade therein (49). The Little Divergence refers to the observation that England, Netherlands, and Belgium experienced faster economic growth than Southern European countries (Italy, Spain, and Portugal) during the centuries leading to the Industrial Revolution. A central explanation for this divergence is the rise of Atlantic trade starting the 16th century. Atlantic trade led to larger direct economic gains and shifted political power towards commercial interests. As Acemoglu et al. argue (49), the latter was not the case in countries with strong absolutist powers, which is why Spain and Portugal profited less from Atlantic trade than England and the Netherlands.

Our regional GDP per capita estimates reproduce these observations (Fig. 3 A-D). While Lombardy was one of the richest regions in Europe up to 1500, with an estimated GDP per capita of around 3,000 2011$, Amsterdam and London experienced higher economic growth in the following centuries. In 1800 Amsterdam and London were among the richest regions in Europe (Fig. 3 A).

To investigate the within-country variation of the Little Divergence we generate population-weighted deciles of GDP per capita for the North (England, Netherlands, Belgium) and the South (Italy, Spain, Portugal). We use the number of births and deaths of famous individuals in a location as population proxies (SI Appendix section 3.3). Our estimates show that the North experienced sustained economic growth between 1300 and 1800, while the South stagnated. Also, we find that, in 1300, the bottom 10$^{th}$ percentile of the South has been as rich as the top 90$^{th}$ percentile of the North. In 1800, the opposite holds: The bottom 10$^{th}$ percentile of the North exhibits a similar income level as the 90$^{th}$ percentile of the South (Fig. 3 B).

We show that Atlantic ports were a significant driver of this development. In line with results by Acemoglu et al. (49), we find that countries with Atlantic ports (UK, NLD, FRA, ESP, PRT) experienced more rapid growth between 1300 and 1850 than other European countries (Fig. 3 C). Moreover, we find that this development is to a large extent driven by regions with Atlantic ports in the United Kingdom and the Netherlands (Fig. 3 D). Their average GDP per capita increased fivefold between 1300 and 1850, from 1,200 to 6,000 USD. In contrast, regions with Atlantic ports in France, Portugal, and Spain, and regions with Mediterranean ports did not experience sustained economic growth during the same period. This supports Acemoglu et al.'s (49) findings using city population as a proxy for regional economic development (SI Appendix section 5.2).

Second, we externally validate our estimates by showing they correlate with four known proxies of economic development: (a) urbanization rates between 1500 and 1950 (54), (b) average body height in the early and late 18$^{th}$ century (55), (c) a composite indicator of well-being in 1850 published by the OECD (56), and (d) city-level church building activity in cubic meters between 1300 and 1450 in Italy, France, Switzerland, the Low Countries, and Great Britain (57). We measure urbanization as the share of urban population (54) relative to total population according to the Maddison project (27, 28). Indeed, urbanization is a frequently employed proxy of pre-industrial living standards and prosperity (49, 99, 100), as is body height (101–103). The OECD well-being indicator aggregates information on GDP per capita, wages, life expectancy, income inequality, years of education, homicide rates, and body height (56). And church building activity is associated with income levels because such projects have been major long-term investments, requiring a positive outlook on the future and the technological advances necessary for such endeavors. In all four cases we find our estimates correlate with these proxy measures (Fig. 3 E-H). We also find these correlations are very similar for labeled and unlabeled observations, alleviating some of the concerns with respect to the generalizability of our results (SI Appendix section 5.3).

Additionally, we explore whether our estimates can recreate patterns of regional development in German regions after the French Revolution as described by Acemoglu and coauthors (104). They find that German regions occupied by the French revolutionary armies, who induced radical institutional changes, experienced larger economic growth (proxied using urbanization rates) in the second half of the 19$^{th}$ century than other German regions. We replicate their descriptive plots with our regional estimates of GDP per capita and find highly similar patterns (SI Appendix section 5.4).

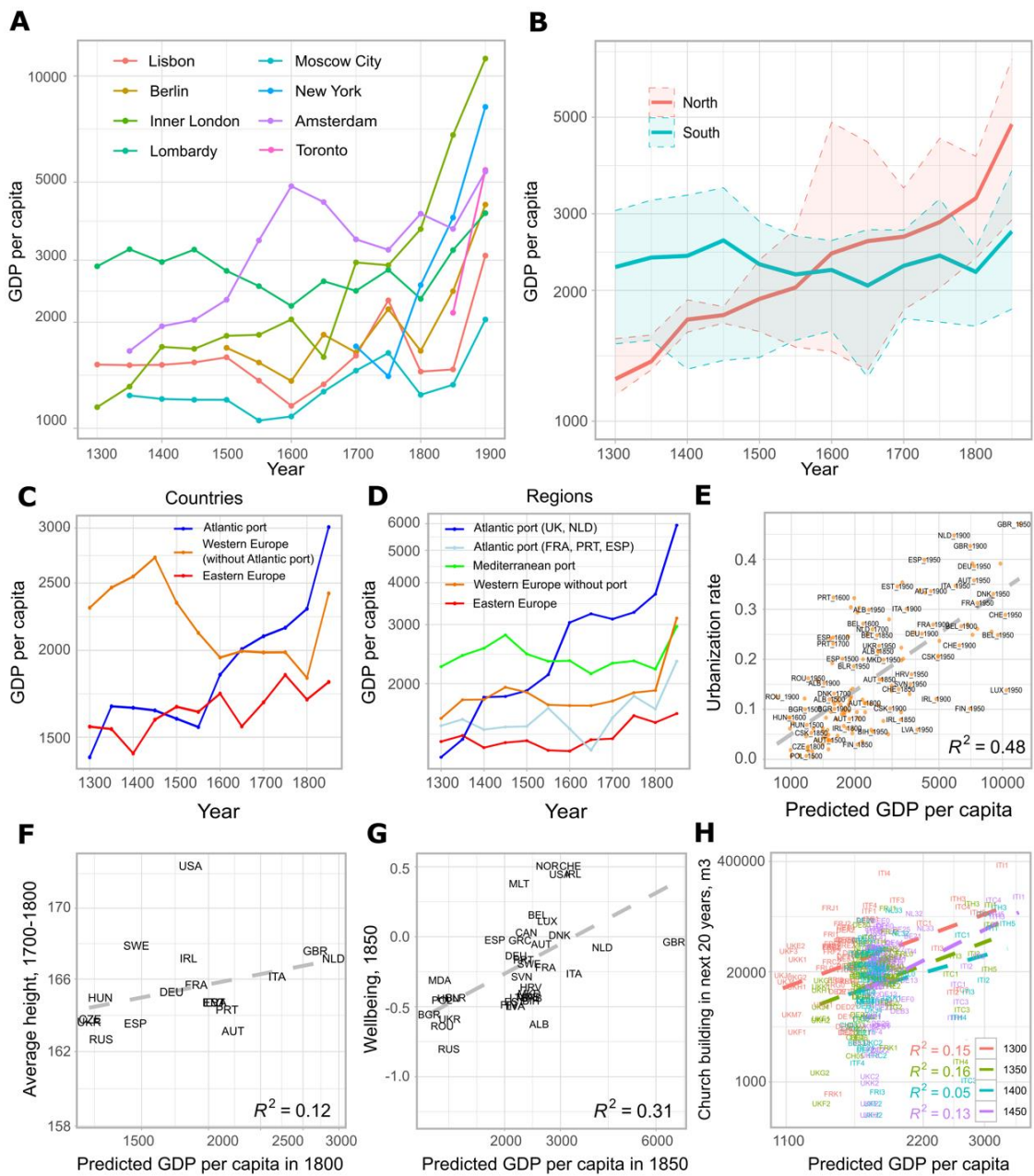

**Figure 3. External model validation.** (**A**) Economic development in selected European and North American regions and cities between 1300 and 1900. (**B**) Little Divergence: England, Netherlands, and Belgium (North) experienced sustained economic growth prior to the Industrial Revolution, while Italy, Spain, and Portugal (South) did not. Displayed are the population-weighted 90th and 10th percentiles, and the mean of the respective GDP per capita. (**C**) Economic development in countries with Atlantic ports, other Western European countries, and Eastern European countries (**D**) Economic development in regions with Atlantic ports, Mediterranean ports, and without a port, showing that Atlantic trade is a relevant driver of the Little Divergence. (**E**) Correlation of predicted GDP per capita with urbanization rates between 1500 and 1950. (**F**) Correlation of predicted GDP per capita with average body height in the 18th century. (**G**) Correlation of predicted GDP per capita with an indicator of wellbeing in 1850 published by the OECD. (**H**) Correlation of predicted GDP per capita with city-level church building activity in the 14th and 15th century.

**Unpacking the evolution of prosperity in Europe and North America**

We now use our estimates to explore some additional stylized facts. On the level of countries, our dataset provides several GDP per capita time series which were yet unavailable, such as Portugal prior to 1530, South Italy prior to 1861, Switzerland prior to 1850, Russia prior to 1885, Austria prior to 1820, and many more. Also, our estimates differentiate the British Isles countries prior to 1700, showing that England was the richest among them after 1400.

Figs. 4 A-C show the evolution of country-level economic development in Europe and North America between 1300 and 1900. In 1300, income levels have been highest in Northern Italy (Fig. 4 A). While the Netherlands and Belgium were among the richest economies in 1600 (Fig. 4 B), we find the United Kingdom and the United States to exhibit the highest income levels in 1900 (Fig. 4 C).

Regional estimates of GDP per capita are even scarcer in published resources. Our dataset enables the investigation of economic development in Europe and North America on a regional level (Figs. 4 D-F). The overall findings are in line with the country-level estimates: Northern Italy became gradually less rich relative to other economies, while the Low Countries and the UK grew sharply. Regional estimates, however, provide more nuance. While the GDP per capita level in Spain was similar to France or England in 1600, we estimate income levels for Madrid (~2,600 USD) to be significantly higher than in London (~2,000 USD) or Paris (~1,800 USD), and even slightly higher than in regions in Northern Italy. Also, we find income levels in Amsterdam in 1600 (~4,900 USD) to be more than 30 percent above other parts of the Netherlands such as Rotterdam (~3,500 USD) or Utrecht (~2,200 USD). In 1900, income levels are more similar across Europe, with Great Britain topping the European charts. The richest cities back then, however, are found in the United States: According to our estimates San Jose and Los Angeles (~13,000 USD) had higher income levels in 1900 than Inner London (~11,200 USD).

We demonstrate three use cases of our data.

First, we know from the Maddison project that Germany was one of the richest economies in Europe in 1500, prior to the Protestant Reformation. But which were the richest regions in Germany back then? Our estimates show that Nuremberg was the region within Germany with the highest GDP per capita in 1500 (Fig. 4 G). The city's prominent position is in line with historical research describing Nuremberg in the 16$^{th}$ century as a renaissance city and cultural and economic center (105, 106). German income levels then fell between 1500 and 1600 on average by 29.6 percent. Nuremberg experienced a similar economic decline, according to our estimates. In contrast, regions that were relatively rich in 1500 but did not experience such a significant decline in the 16$^{th}$ century are Swabia (with its capital Augsburg) and Rheinhessen-Pfalz (incl. the cities Frankenthal and Kaiserslautern). One possible explanation is that cities in Swabia and Rheinhessen-Pfalz adopted Protestantism relatively early in the Reformation, and Protestantism has (ever since Max Weber) been connected to positive economic outcomes (107, 108). The link between Protestantism and economic prosperity, however, is not unquestioned. A empirical analysis of 272 cities in the Holy Roman Empire shows that there is no association between Protestantism and population growth (109). Here, we find that Protestant regions such as Nuremberg, Swabia and Rheinhessen-Pfalz were among the richest regions in 1500 and the latter two experienced less economic decline in terms of income per capita over the course of the 16$^{th}$ century than other German regions.

Second, we can use our estimates to explore the history of Charleston, South Carolina. Charleston emerged as a commercial hub and major city between 1720 and 1730 (110). We find it to be one of the richest metropolitan areas in North America in 1750 (Fig. 4 H). After the American Revolution, Charleston was the largest city in the South, continuing to be a center for slave trade (111). Our estimates reflect that since Charleston did not develop as positively as other cities during the antebellum era (Fig. 4 H).

Third, we find that the GDP per capita of Lisbon declined sharply after 1750 (Fig. 3 A). This observation coincides with the disastrous earthquake that hit Portugal's South in 1755 and had severe economic consequences (112). The Maddison project estimates the GDP per capita of Portugal fell by 33.2% between 1750 and 1800, and we estimate Lisbon's GDP per capita fell by 37.2% in this period. In contrast, we find that the GDP per capita of regions in Portugal that were not as affected by the earthquake even developed positively: Income per capita grew by 6.6% in Northern Portugal and by 9.5% in the region Alentejo.

**Feature importance**

Finally, we explore the importance of the features selected by our model before providing additional evidence about the robustness of our results. We unpack feature importance using Shapley values. Shapley values originate from game theory (113) and are frequently applied in machine learning to interpret predictions (114, 115). These are defined as the average marginal effect of including a certain feature over all possible feature combinations (*Materials & Methods*).

Figs. 4 I-K show the most relevant features in 1300, 1600, and 1900, respectively. In 1300, the dummy variable for Eastern Europe is the most relevant feature, correlating negatively with GDP per capita. Looking at interpretable features derived from biographies, we find that being a place of deaths for famous lawyers and painters, and a place of birth for famous politicians are among the most relevant positive predictors of GDP per capita in 1300 (Fig. 4 I). In 1600, we find that the GDP per capita in the previous period is the most relevant feature in predicting GDP per capita levels. Also, the number of deceased and immigrant priests correlates negatively with income levels, while the number of deceased, born, and immigrant painters correlates positively (Fig. 4 J) with GDP per capita. We also find some SVD factors to be relevant features in 1600, such as the third factor describing the geography of famous births and the fourth factor describing the geography of famous deaths (SI Appendix section 4.2). These abstract factors, however, lack a direct interpretability compared to the number of births and deaths of individuals with a given occupation. In 1900, next to the initial income level, the diversity of occupations as well as the average age of famous individuals in a location are positive predictors of income levels (Fig. 4 K).

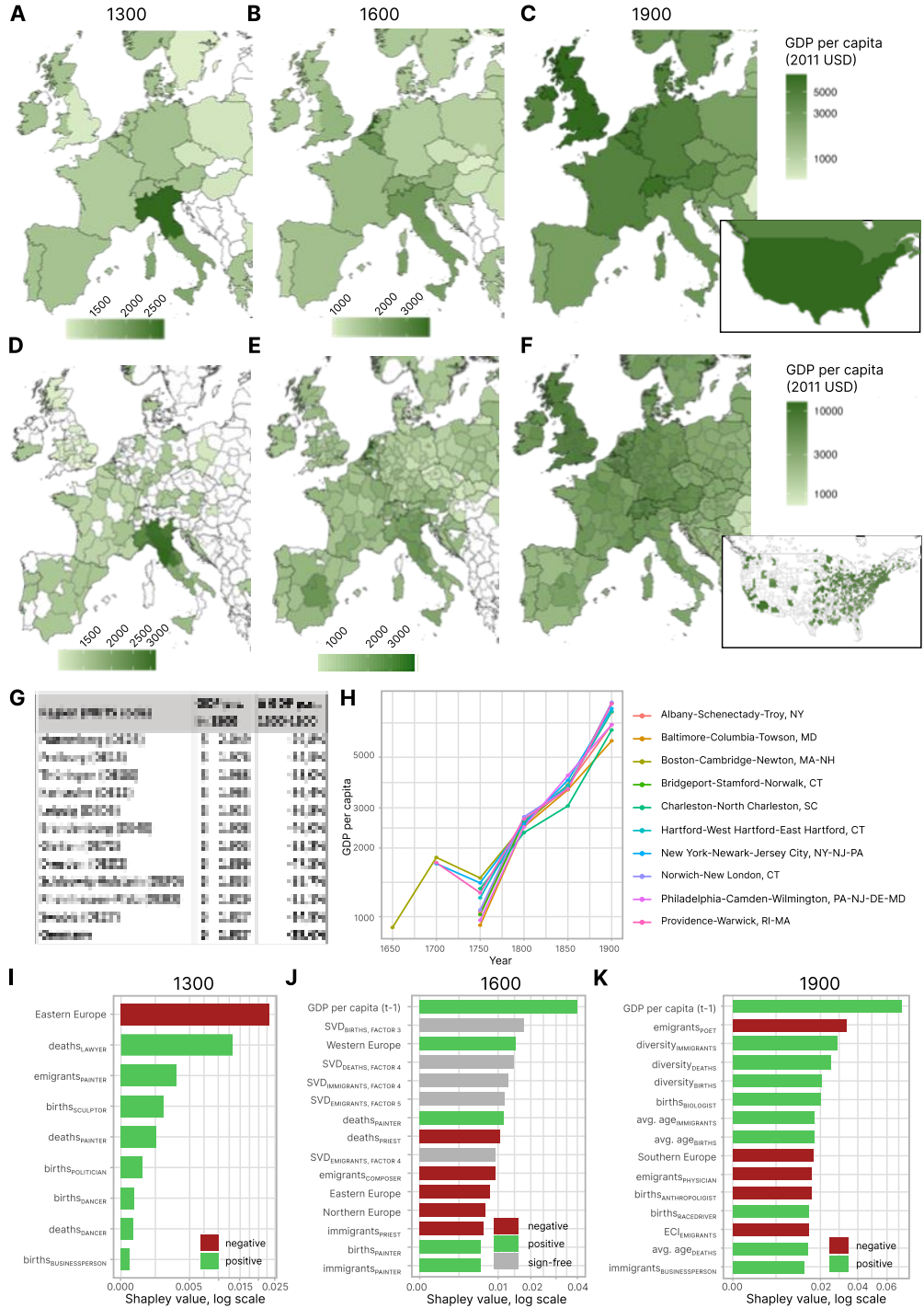

**Figure 4. Evolution of prosperity in Europe and North America.** (**A-C**) Country-level GDP per capita estimates in Europe and North America in (**A**) 1300, (**B**) 1600, and (**C**) 1900. (**D-F**) GDP per capita in European and North America regions and cities in (**D**) 1300, (**E**) 1600, and (**F**) 1900. (**G**) Richest regions in Germany in 1500 and economic growth in the 16th century. (**H**) Economic development of selected metropolitan areas in the United States between 1650 and 1900. (**I-K**) Feature importance measured in Shapley values for (**I**) 1300, (**J**) 1600, and (**K**) 1900.

**Robustness of our estimates**

We check the sensitivity of our results to biases in the data and justify our methodological choices through several robustness checks (SI Appendix section 5.5). First, we investigate how our model performs when we use only data prior to the year 2000, since relatively recent time periods may upward bias our model's performance measures. We find this is not the case. While the $R^2$ is lower, model performance in terms of the mean absolute error improves slightly when we remove data for the year 2000 (SI Appendix section 5.5.1). Second, we investigate whether the English bias in Wikipedia significantly affects our estimates. For this purpose, we compare our results to those obtained using only English Wikipedia pages or only non-English Wikipedia pages. All three samples yield highly similar results, even for regions in English-speaking countries, indicating that this data limitation is not driving our estimates (SI Appendix section 5.5.2). Third, we provide model performance results for other thresholds of assigning biographies to time periods. We find that other thresholds yield similar, but slightly worse results than using 150 years (SI Appendix section 5.5.3). Fourth, we linearize our features before fitting our regression models. We use logarithms in our main results but provide robustness checks using the inverse hyperbolic sine function. We find that both scaling functions yield similar results (SI Appendix section 5.5.4). Fifth, we test whether backward feature selection performs better than EN regression models. We find that backward feature selection performs significantly worse (SI Appendix section 5.5.5). Sixth, we test whether our model is sensitive to the use us of HPI when deriving features from the biographies of historical figures, and find that removing the HPI slightly decreases model performance (SI Appendix section 5.5.6). Seventh, we test to what extent the dummies for supranational regions are driving our results. Removing them from the elastic net model yields only slightly worse results (SI Appendix section 5.5.7). Lastly, we investigate whether we can predict GDP per capita growth rates instead of levels following the same methodology. Here, we do not find a significant improvement compared to the baseline, a fact that could come from the significantly lower number of observations we have for growth (we need two observations for each growth number, meaning that we have only 455 ground truth observations for growth compared to over 1,300 for income levels) (SI Appendix section 5.5.8).

## Discussion

Despite significant efforts to collect data on historical income levels (27–29, 58–65), our understanding of long-term economic development remains limited. Here, we explored whether data on the biographies of historical figures can be used to create models of historical GDP per capita levels for countries and regions in Europe and North America for the past 700 years and estimate their historical GDP per capita.

This method is, however, not without limitations. First, our data on GDP per capita levels going back centuries must be understood as estimates of estimates. That is, the "ground-truth" data we use to generate out-of-sample estimates are already estimates. This induces a level of uncertainty that needs to be considered when using our data and method. Second, data from Wikipedia can lack accuracy and is known to be subject to biases (77). Data from Wikidata, the structured database related to Wikipedia, can lack information (31). We know, for instance, that Yury of Moscow appears in Wikipedia as born in Moscow in 1281, but Wikidata does not report that information. Despite efforts to cross-validate Wikipedia and Wikidata (31, 32), future research can improve its accuracy. Similarly, we do not observe all places an individual lived in, but only the places of birth and death. To show that our estimates are not affected by language biases, we provide several robustness checks and are careful to not extend our estimates to Africa, Asia or South America. Third, we provide results using elastic net models since they are efficient in selecting features and preventing overfitting, but future research may come up with better models and methods (e.g. backcasting). Lastly, countries and regions for which source data is available are not perfectly representative of locations without available source data. Indeed, countries and regions with source data tend to have a higher GDP per capita in 2000 and a higher number of famous individuals than countries and regions without source data. Still, we find that the correlation between our estimates and proxies of economic development is comparable for labeled and unlabeled observations, which alleviates some of the concerns with respect to the generalizability of our results (SI Appendix section 5.3).

Together, this paper introduces a new method for the generation of historical GDP per capita estimates with encouraging results and showcases the use of structured historical data for the estimation of long-term economic time series. Specifically, our findings validate the use of fine-grained biographical data as a method to produce historical GDP per capita estimates. We hope future research can build upon this idea to further improve our understanding of economic development. We publish our estimates with confidence intervals together with all collected source data and the code to replicate our results. This dataset does not only allow for investigating 700 years of cross-country differences in economic development, but also for comparing the development of different regions in Europe (Milan, Montpellier, Paris, London, etc.) with metropolitan areas in North America (New York, Boston, Toronto etc.).

## Materials and Methods

**Historical Popularity Index**. We take the historical popularity of individuals in our dataset into account when defining features. We reconstruct a version of the Historical Popularity Index (*HPI*) introduced in the Pantheon database (32) with available data. Specifically, an individual's *HPI* is proportional to the number of Wikipedia page views ($V$), the number of language editions ($L$) and age ($A$, i.e. 2023 minus year of birth):

$$HPI = \begin{cases} log_{10}(V) + ln(L) + log_4(A) & if\ A \geq 70 \\ log_{10}(V) + ln(L) + log_4(A) - \dfrac{70-A}{7} & if\ A < 70 \end{cases}$$

This measure of historical popularity is strongly correlated with the *HPI* in the Pantheon dataset (which also includes information on the entropy of the distribution of pageviews across languages and uses information on pageviews in non-English editions of Wikipedia) ($R^2 = 0.76$, SI Appendix section 3.2).

**Economic Complexity**. To calculate economic complexity, we create binary adjacency matrices $M_{ik,t}$ which indicate whether a location is specialized in an occupation based on measures known as the Revealed Comparative Advantage or Location Quotient:

$$M_{ik,t} = \begin{cases} 1 & if\ \dfrac{N_{ik,t}/N_{i,t}}{N_{k,t}/N_t} \geq 1 \\ 0 & otherwise \end{cases},$$

where $N_{ik,t}$ denotes the number of famous individuals in location *i* with occupation *k*, weighted by their $HPI$. Then, the economic complexity index ($ECI$) is defined as the result of an iterative mapping, defining a location's complexity as the average complexity of the occupations it is specialized in:

$$ECI_i = \frac{1}{M_i} \sum_k M_{ik} PCI_k$$

$$PCI_k = \frac{1}{M_k} \sum_i M_{ik} ECI_i \quad .$$

We compute separate $ECI$s for births, deaths, immigrants, and emigrants (SI Appendix section 4.1).

**Singular Value Decomposition**. Singular Value Decomposition (SVD) is a dimensionality reduction technique which retrieves factors from a rectangular matrix that best explain its structure. Here, we collect our data in adjacency matrices $N_{ik,t}^j$ describing the (HPI-weighted) number of births, deaths, immigrants, or emigrants in a certain location with a certain occupation. Index $i$ denotes the location, $k$ denotes the occupation and $j$ differentiates between births, deaths, immigrants, and emigrants.

Mathematically, SVD decomposes matrix $N$ (technically, we use its logarithm) into

$$N = U \times S \times V^T ,$$

where $U$ and $V^T$ are unitary matrices collecting orthonormalized eigenvectors describing locations and occupations, respectively, and $S$ is a diagonal matrix collecting the singular values (116). We include the first five eigenvectors in $U$ for births, deaths, immigrants, and emigrants as candidate features, i.e. twenty potential features per period (SI Appendix section 4.2).

**Elastic Net**. We use elastic net (EN) regression models to perform feature selection and generate out-of-sample estimates. EN does not simply minimize the sum of squared residuals like an OLS regression would do, but also penalizes for the $\ell^1$ and $\ell^2$ norms of the coefficients, effectively performing feature selection and reducing the risk of overfitting. Mathematically, the EN estimator $\widehat{\boldsymbol{\beta}}$ minimizes the following function $L$ for given parameters $\alpha$ and $\lambda$:

$$L(\alpha, \lambda, \boldsymbol{\beta}) = \|\boldsymbol{y} - \boldsymbol{X\beta}\|^2 + \lambda(\alpha\|\boldsymbol{\beta}\|_1 + (1-\alpha)\|\boldsymbol{\beta}\|_2^2) ,$$

where $\boldsymbol{y}$ is the log of GDP per capita (base 10) and $\boldsymbol{X}$ is a vector of features. Note that the EN collapses to a LASSO (least absolute shrinkage and selection operator) if $\alpha = 0$ and to a ridge regression if $\alpha = 1$. The parameter $\lambda$ controls the extent of the penalty. We find optimal values for $\alpha$ and $\lambda$ using k-fold cross-validation ($k = 10$), minimizing the prediction error. Parameter values and selected features for each period are provided in SI Appendix section 5.1.

The models account for persistence in income levels by including the GDP per capita from the end of the previous historical period as a potential feature. The latter variable is not available for all locations and all time periods, so we use the following hierarchical approach (analogous to the baseline model). If available, we use the GDP per capita at the end of the previous period from source data. If that is not available, we use the estimates of the EN model in the previous historical period. For regions with unavailable source data or EN model estimates for the previous period, we use instead the data or model estimates of the country that region is in. If none of the above is available, we use the average of the supranational region at the end of the previous period as initial GDP per capita.

**Model performance.** We test how well our model performs on out-of-sample data using 500 randomly drawn, independent test sets. Specifically, one iteration (out of 500) of assessing the model's performance consists of, first, randomly selecting 20% of countries. For the model performance to be accurate and unbiased, it is crucial to make sure the test set (the withheld 20% of countries) is independent of the choice of hyperparameters. Hence, we now use the remaining 80% of countries to tune the hyperparameters of the EN model ($\alpha$ and $\lambda$).

For tuning $\alpha$ and $\lambda$, we use 10-fold cross-validation. That is, the sample of 80% of countries is split into 10 subsamples. Then, we find hyperparameters by, iteratively, leaving one of those subsamples out (validation sets) and using the remaining 9 subsamples as training sets. The optimal hyperparameters are the averages over these 10 iterations. Next, we use this model (trained on 80% of the countries) to predict the GDP per capita of the remaining 20% of countries, which the model has not encountered yet, and compare our estimates with the respective source data (using R-squared and mean absolute error). We compute the R-squared using the estimates for the log of GDP per capita, and use the exponentiated estimates for computing the mean absolute error. This procedure is repeated 500 times, eventually yielding Fig. 2 C-D.

**Shapley values**. Shapley value $\phi_i$ is defined as the average marginal effect of including feature $i$ in the model for all possible feature combinations $S$:

$$\phi_i = \sum_{S \subseteq F_{-i}} \frac{|S|!\,(|F| - |S| - 1)!}{|F|!} [f_{S \cup i}(x_{S \cup i}) - f_S(x_S)] ,$$

where $F$ denotes the set of all model features.

**Data and code availability.** We publish our out-of-sample estimates together with the collected source data on countries (27, 28) and regions (29, 58–65) in a comprehensive dataset comprising 5,700 observations (1,336 source data observations, and 4,364 out-of-sample estimates). For the out-of-sample estimates, we provide 90 percent confidence intervals. Also, we publish the code to ensure reproducibility of our results. Data and code are available at https://github.com/philmkoch/historicalGDPpc.

**Supporting Information.** Supporting Information is available at the article's PNAS page: https://www.pnas.org/doi/10.1073/pnas.2402060121.

## Acknowledgments

We acknowledge the support of the Agence Nationale de la Recherche grant number ANR-19-P3IA-0004, the 101086712-LearnData-HORIZON-WIDERA-2022-TALENTS-01 financed by European Research Executive Agency (REA), IAST funding from the French National Research Agency (ANR) under grant ANR-17-EURE-0010 (Investissements d'Avenir program), and the European Lighthouse of AI for Sustainability [grant number 101120237-HORIZON-CL4-2022-HUMAN-02].

**Author Contributions:** All authors designed research and wrote the paper, P.K. performed research and analyzed data, C.A.H. supervised the project.